\documentclass[twocolumn,showkeys,preprintnumbers,amsmath,amssymb]{revtex4}

\usepackage{graphicx}
\usepackage{dcolumn}
\usepackage{bm}
\usepackage{hyphenat}
\begin{document}

\preprint{}

\title{Change in activity character of coronae of low-mass stars of various spectral types}

\author{B.A. Nizamov}
 \altaffiliation[Physical Faculty of Lomonosov Moscow State University]{}
 \email{nizamov@physics.msu.ru}
\author{M.M. Katsova}%
 \email{maria@sai.msu.ru}
\affiliation{%
Sternberg State Astronomical Institute of Lomonosov Moscow State University
}%

\author{M.A.Livshits}
 \email{maliv@mail.ru}
\affiliation{
Pushkov Institute of Terrestrial Magnetism, Ionosphere and Radiowaves Propagation of Russian Academy of Sciences (IZMIRAN)
}%


\begin{abstract}
We study the dependence of the coronal activity index on the star's rotation rate. This question was considered earlier for 824 late-type stars on the basis of a consolidated catalogue of the soft X-ray fluxes. We carry out a more refined analysis separately for G, K and M dwarfs. They distinctively exhibit two modes of activity. The first one is the saturation mode, it is characteristic of young stars and is practically not related to their rotation. The second one refers to the solar-type activity the level of which strongly depends on the rotation period. We show that the transition from one mode to another takes place at the rotation periods of 1.1, 3.3 and 7.2 days for the stars of spectral types G2, K4 and M3 respectively. In the light of the discovery of superflares on G and K stars on the {\itshape Kepler} spacecraft there arises a question of how these objects differ from other active late-type stars. We analyse the location of superflare stars relative to the stars observed by {\itshape Kepler} on the ``amplitude of rotational brightness modulation (ARM) -- rotation period'' diagram. The value of ARM reflects the relative spots area on a star and characterises the activity level in the whole atmosphere. It is shown that G and K superflare stars are basically fast rotating young objects, but some of them belong to the stars with the solar type of activity.
\end{abstract}

\maketitle

\section{\label{sec:level1}Introduction}

The modern solar-stellar physics is characterised by obtaining observational data not on single objects, but on large numbers of stars. First of all this refers to the study of the phenomena in stellar photospheres by spacecrafts such as {\itshape CoRoT}, {\itshape Kepler} etc., which provide us with the information on tens and hundreds thousand objects. Some effects, such as bimodality of the rotation velocities distribution of stars, are now confidently revealed. This effect is distinctly manifested in M dwarfs which can be subdivided to the stars with the activity saturation and more quiet stars with the activity level close to that of the Sun (McQuillan et~al.~2014). Obviously, the main factor determining the level of activity is the rotation velocity. This applies to the activity in the chromosphere and the corona. Active processes develop approximately in the same way on all the low-mass stars possessing subsurface convective zones. However the dependence of the level and character of the activity on the spectral type is pronounced somewhat weakly in F, G, K and M dwarfs so that only recently has it become possible to detect this dependence. Here we consider this question in regard to the X-ray emission of stars.

Massive exploration of the stellar coronal activity has become possible due to the creation of the catalogue of active F-M stars constructed on the basis of X-ray observations onboard the {\itshape Einstein}, {\itshape ROSAT}, {\itshape XMM-Newton} spacecrafts by Wright et~al.~(2011). They derived the dependence of the coronal activity index $R_\mathrm{X} = \lg L_\mathrm{X}/L_\mathrm{bol}$ on the rotation velocity and reliably showed that there exist two groups of stars. First, a large group with the activity saturation where $R_\mathrm{X}$ is close to $10^{-3}$ and practically does not depend on the rotation period. This group includes fast rotating stars. Second, a group of less active stars which demonstrate clear dependence of $R_\mathrm{X}$ on the rotation period. We will call the activity of these stars ``solar-type activity''. This is the magnetic activity which is characterised by the formation of spots, active regions and flares. We assume that regular cycles form within this type of activity. In more detail the evolution of the activity is considered by Katsova \& Livshits (2011) and Katsova et~al.~(2015). The stars in the second group obey Skumanich law $v \sim \sqrt{t}$, where $v$ is the rotation velocity, $t$ is the age of the star. Thus, the decrease of $R_\mathrm{X}$ with period (i. e. increase of $R_\mathrm{X}$ with the rotation velocity) can be related to the increase of the age, the fact that underlies gyrochronology --- the estimation of the age on the basis of the activity level (Mamajek \& Hillenbrand 2008). Based on the dependence of the chromospheric and coronal activity indices on the rotation period, this method has been developed for stars of different ages in the lower part of the main sequence without a separate analysis of the influence of the difference in spectral types. First evidence of the difference of the surface activity level among dwarfs of different spectral types are found by Messina et~al.~(2003) who analyse the rotational modulation of the stellar optical radiation related to the spots and consider the radiation of the coronae. The subsequent comparison of the chromospheric and coronal activity of late-type stars showed that the method of gyrochronology cannot be based on a single parameter --- such as the rotation period --- but instead should take into account the difference in spectral types (Katsova \& Livshits 2011).

Recently there appeared a new approach to the analysis of X-ray data on stellar coronae. The catalogue by Wright et~al.~(2011) with minor refinements was used by Reiners et~al.~(2014). They looked for the dependence of $L_\mathrm{X} / L_\mathrm{bol}$ on the combination of the rotation period and radius of the form $R^\alpha P^\beta$ and found that for the stars with solar-type activity the data are best described by the model with $\alpha=-4$, $\beta=-2$. Taking approximately $T_\mathrm{eff} \sim R^{1/2}$ for G-M dwarfs the bolometric luminosity scales with the radius approximately as $L_\mathrm{bol} \sim R^4$. Thus, the aforementioned dependency reduces to the law $L_\mathrm{X} \sim P^{-2}$, earlier obtained by Pizzolato et~al.~(2003). The results of Reiners et~al.~(2014) has become an argument in favour of that the axial rotation rate of a star is indeed the main factor determining the activity level of different layers of the atmosphere.

However in the aforementioned papers (Wright et~al.~2011; Reiners et~al.~2014) late-type stars were considered together, without subdivision in to spectral types. The main goal of our work is to extract the information on the dependence of the activity on the spectral type from the same data. We carry out an investigation analogous to that of Reiners et~al.~(2014), but separately for G, K and M stars. Further we discuss the effect of bimodality of the rotation periods distribution and its relation to the features of the stars where superflares were detected (the properties of these stars are studied by Shibayama et~al.~2013; Notsu et~al.~2015). In conclusion, we discuss our results from the point of view of changes in physical conditions in the corona which accompany the transition from a fast rotating star to slower rotating ones.

\section{Relation between X-ray luminosity of late-type stars and their rotation period.}\label{main}
Our goal is to reveal the relation between X-ray activity on the one hand and radius and rotation period on the other. Following Reiners et~al.~(2014) we assume that this relation is of the (quite general) form $L_\mathrm{X}/L_\mathrm{bol} = k R^\alpha P^\beta$ and in logarithmic scale becomes linear:
\begin{equation}
	\lg \frac{L_\mathrm{X}}{L_\mathrm{bol}} = \lg k + \alpha\lg R + \beta\lg P. \label{nonsat}
\end{equation}
This is an equation of a plane in the coordinates $\lg R$, $\lg P$. We will turn to such a representation later, but to carry out the computations it will be more convenient to move to the relation between $\lg L_\mathrm{X}/L_\mathrm{bol}$ and only one (not two) independent variable. Let us denote the right hand side of the equation (\ref{nonsat}) $x$. A straight line $\lg L_\mathrm{X}/L_\mathrm{bol} = x$ will describe the X-ray luminosity of the stars with the solar-type activity. For the stars with the saturation of activity we will approximate the data by the relation of the form $\lg L_\mathrm{X}/L_\mathrm{bol} = cx + d$. Thus, the relation between $L_\mathrm{X}/L_\mathrm{bol}$ and $R^\alpha P^\beta$ is described by two straight lines in logarithmic scale:
\begin{equation}
\begin{cases}
	y=\lg\frac{L_\mathrm{X}}{L_\mathrm{bol}} = x, & x \ge x_0,\\
	y=\lg\frac{L_\mathrm{X}}{L_\mathrm{bol}} = cx+d, & x < x_0.
\end{cases}
\end{equation}
The point $x_0$ of transition from one mode to another is a parameter of the model. In order to apply the standard least squares we have to represent the model relation as a single function. We do this as follows:
\begin{equation}
	y = f(x) = \sigma(x) x + (1-\sigma(x))(cx + d),
\end{equation}
where $\sigma(x)$ is a weight function which should take the value 1 if \mbox{$x \ge x_0$} and 0 if \mbox{$x < x_0$}. Since we need a differentiable function, we choose not the Heaviside step function, but its differentiable ``analogue'' $\sigma(x) = (1+e^{-100(x-x_0)})^{-1}$.

Now that we have the values $\lg L_\mathrm{X}/L_\mathrm{bol}$ (i. e. $y$) as well as $\lg P$, $\lg R$ for each star in the dataset, we can compute the values of $x$ and find the parameters $k$, $\alpha$, $\beta$, $c$, $d$, $x_0$ by means of least squares applying, e. g., gradient descent. The function to minimise is the sum of the squares of the residuals:
\begin{equation}
	S = \sum\left[\lg\left(\frac{L_\mathrm{X}}{L_\mathrm{bol}}\right)_i - f(x_i)\right]^2.
\end{equation}

Let us make a remark on the method we applied. The overall result of the calculation is the model relation between $\lg L_\mathrm{X}/L_\mathrm{bol}$ and {\itshape one} variable $x$. Since there are two variables $\lg R$ and $\lg P$ accumulated in $x$, it might seem that there should be a degeneracy between the parameters $\alpha$, $\beta$ and $k$. To show that this is not the case let us give a slightly different interpretation of the optimisation method applied. Consider the branch of the sought-for relation which corresponds to the solar-type activity. The initial data of the problem are the values $\lg L_\mathrm{X}/L_\mathrm{bol}$, $\lg R$ and $\lg P$, i. e. the points in a three-dimensional space. The combination of the two latter into a single variable $x$ means that all the points are projected onto a plain $\Pi$ containing the $\lg L_\mathrm{X}/L_\mathrm{bol}$ axis. In this plane we find a position of a straight line so that the scatter of the points around this line is minimal. The position of the plane $\Pi$ is determined only by the relation $\alpha / \beta$, which implies the uncertainty in the individual parameters. But the position of the fitting line does depend on individual values $\alpha, \beta$ and this removes the uncertainty.

\begin{figure}
\includegraphics[scale=0.8]{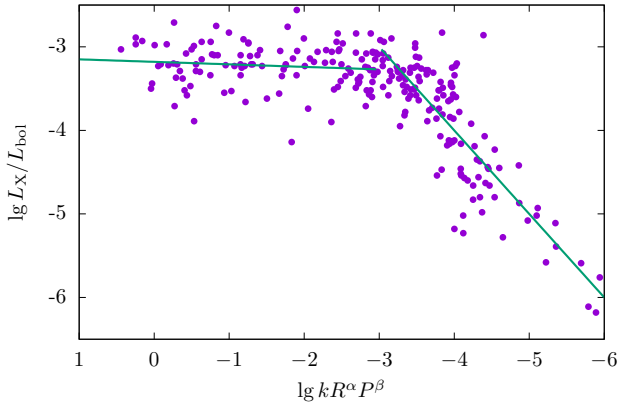}
\caption{\label{fig1} The coronal activity index vs. $\lg k R^\alpha P^\beta$ for K stars.}
\end{figure}

In Fig.~\ref{fig1} we show the K stars from the dataset as well as the obtained model relation $\lg\frac{L_\mathrm{X}}{L_\mathrm{bol}}(\lg R, \lg P)$ in both modes (the figure is analogous to Fig.~3 {\itshape right} from Reiners et~al.~2014, but the ours is for the K stars only). The transition from one mode of activity to another formally takes place abruptly which is somewhat unphysical, but this is probably caused by the overall scatter of the data and errors in the model parameters determination.

\begin{figure*}
\includegraphics{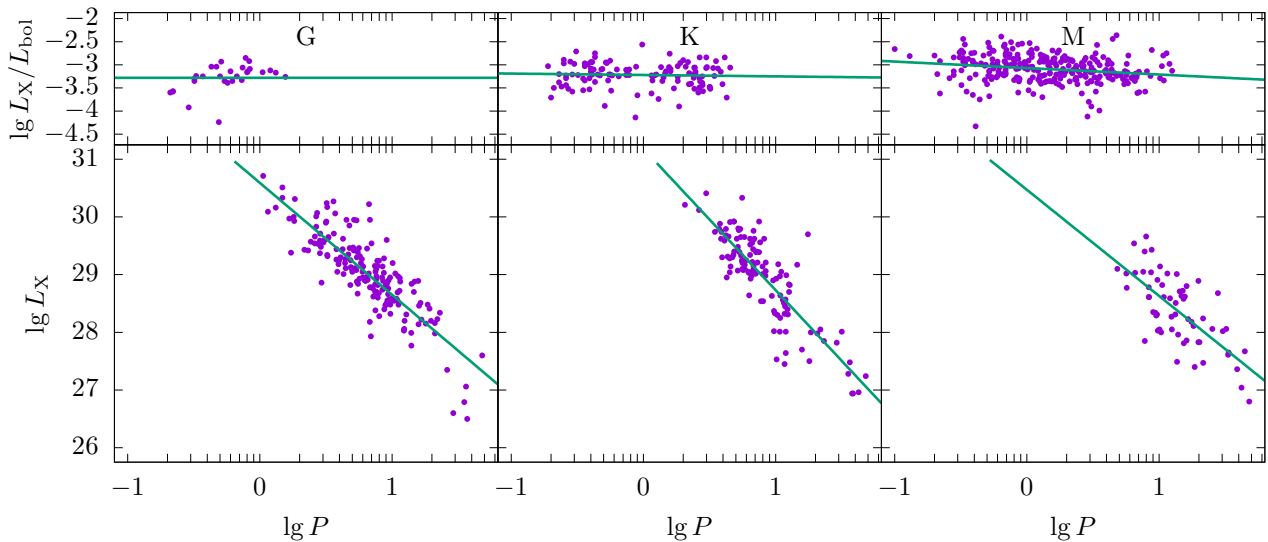}
\caption{\label{fig2}Coronal activity index vs. logarithm of rotation period for the stars with the saturated activity (upper row) and X-ray luminosity vs. logarithm of rotation period for the stars with the solar-type activity (lower row). See the note on the upper left plot in the text.}
\end{figure*}

The values of the optimisation parameters for G, K and M stars are given in Table~\ref{tab1} as well as the parameter values for the whole dataset of 824 stars.
\begin{table*}
\caption{\label{tab1}Parameters of the model relation $\lg\frac{L_\mathrm{X}}{L_\mathrm{bol}}(\lg R, \lg P)$ which minimise the scatter of the dataset (shown in the last column).  }
\begin{ruledtabular}
\begin{tabular}{lcccccccc}
Spectral type & Number & $\lg k$ & $\alpha$ & $\beta$ & $c$ & $d$ & $\lg x_0$ & $\sigma$ \\
\hline
G & 202 & $-3.01 \pm 0.09 $ & $-4.66 \pm 0.59$ & $-1.91 \pm 0.12$ & $-0.21 \pm 0.09$ & $-3.77 \pm 0.23$ & $-3.12 \pm 0.10$ & 0.356 \\
K & 242 & $-2.32 \pm 0.25 $ & $-3.48 \pm 0.73$ & $-2.41 \pm 0.21$ & $0.03 \pm 0.02$ & $-3.18 \pm 0.06$ & $-2.98 \pm 0.26$ & 0.335 \\
M & 344 & $-2.84 \pm 0.63 $ & $-4.06 \pm 0.86$ & $-2.17 \pm 0.57$ & $0.07 \pm 0.05$ & $-3.04 \pm 0.18$ & $-3.02 \pm 0.46$ & 0.308 \\
All & 824 & $-2.85 \pm 0.11$ & $-4.03 \pm 0.29$ & $-2.04 \pm 0.13$ & $0.08 \pm 0.01$ & $-3.05 \pm 0.03$ & $-3.27 \pm 0.15$ & 0.345\\
\end{tabular}
\end{ruledtabular}
\end{table*}
The latter are close to that found by Reiners et~al.~(2014), as expected. Note that we used the catalogue of Wright et~al.~(2011) with no corrections which were made by Reiners et~al.~(2014). The errors of the parameters determination were calculated by bootstrap method (see Appendix). One can see that the parameters for M stars are close to the mean ones. The reason is that it is M stars that prevail in the dataset. Moreover, the scatter of these stars around the model curve is even smaller than for the whole dataset (0.308 vs. 0.345). This is apparently due to the large fraction of M stars with the saturated activity. Such stars have higher X-ray luminosity which is determined more precisely than for the stars with the solar-type activity. It is also seen that the parameters $\alpha$ and $\beta$ for M stars have rather low accuracy because only a small part of these stars is in the mode of the solar-type activity. This is illustrated by Fig.~\ref{fig2} where all the G, K, M stars are separated according to their activity type. The upper row represents the stars with the saturated activity. The lower row shows the stars with the solar-type activity. Note that the $y$ axes units are different in two cases. The straight line in the upper left plot is not a least squares fit because of lack of reliable data points, it only illustrates the average level of $\lg L_\mathrm{X} / L_\mathrm{bol}$ among G stars  with the saturated activity in our set.

The parameters for the K stars noticeably differ from the mean ones. In particular, the parameter $\beta$ does not equal 2 even within the error box. The same applies to the parameter $\alpha$ for G stars which substantially differs from 4.

The straight line which describes the solar-type activity is only determined by the parameters $k$, $\alpha$ and $\beta$. In the dataset used the amount of the stars with this type of activity decreases from G to M type (see Fig.~\ref{fig2}), so that the errors of these parameters determination increase towards the cool stars.

For the parameters $\alpha$, $\beta$ Reiners et~al.~(2014) found the values 4 and 2 respectively, i. e. $L_\mathrm{X}/L_\mathrm{bol} = kR^{-4}P^{-2}$. Taking for the stars with the solar-type activity an approximate relation $T_\mathrm{eff} \propto R^{1/2}$ (see, e. g., Allen~1973), they obtained $L_\mathrm{X}/L_\mathrm{bol} \propto P^{-2} T_\mathrm{eff}^{-4} R^{-2} \propto P^{-2} L_\mathrm{bol}^{-1}$ and consequently $L_\mathrm{X} \propto P^{-2}$, i.~e. the X-ray luminosity of a star in the mode of solar-type activity depends only on the rotation period which implies in turn that the period at which the transition from the saturated activity to the solar-type activity takes place depends solely on the bolometric luminosity.

Now we carry out the analysis separately for three spectral types. First of all note that in order the X-ray luminosity to depend only on the period it is necessary that $\alpha=4$. It is the case within the error boxes for K and M stars and hardly for G stars. If we assume that the relation $L_\mathrm{X} \propto P^{-2}$ does apply, we'll be able to determine, like Reiners et~al.~(2014), the rotation period, where one mode of activity changes to another, as a function of the star's luminosity. Since for different spectral types we have obtained the values of $\alpha$ which are not precisely equal to 4, we determine this period as follows. For the ``break point'' we have an expression $k R^\alpha P^\beta = 10^{x_0}$. Using the relation $L_\mathrm{bol} = L_\odot (R/R_\odot)^4$, let us rewrite this expression in the form $k L_\mathrm{bol}^{\alpha / 4} P^\beta = 10^{x_0}$ (the radii and luminosities are expressed in solar units, the periods in days), or $P = (10^{x_0} / k)^{1/\beta} (L_\mathrm{bol}/L_\odot)^{-\alpha/4\beta}$. In this way follow the relations for G, K and M stars:
\begin{align}
	P_\mathrm{sat\,G} &= 1.14 \left(\frac{L_\mathrm{bol}}{L_\odot}\right)^{-0.61}, \\
	P_\mathrm{sat\,K} &= 1.88 \left(\frac{L_\mathrm{bol}}{L_\odot}\right)^{-0.36}, \\
	P_\mathrm{sat\,M} &= 1.21 \left(\frac{L_\mathrm{bol}}{L_\odot}\right)^{-0.47},
\end{align}
which give for the stars of spectral types G2, K4, M3 the values 1.1, 3.3 and 7.2 days respectively (the luminosities are calculated from the effective temperatures and radii given in Strai\v{z}ys \& Kuriliene~1981). One sees that cool stars remain in the saturated mode until much larger periods than warmer stars. In other words, in the course of braking the saturation mode on an M star lasts much longer than on a G star. Reiners et~al.~(2014) obtain for the whole dataset a relation $P_\mathrm{sat} = 1.6 (L_\mathrm{bol} / L_\odot)^{-0.5}$. Having known the period at which the change of the activity modes occurs, one can plot the relation between $\lg L_\mathrm{X} / L_\mathrm{bol}$ and $\lg P$ for the stars with the saturated activity and the relation between $\lg L_\mathrm{X}$ and $\lg P$ for the stars with the solar-type activity. These plots are shown in Fig.~\ref{fig2}. One can see that among the G stars the overwhelming majority comprises the stars with the solar-type activity. In contrast, the M stars mainly show the activity saturation. Both activity modes are well represented among the K stars.

\section{What stars are superflares detected on?}
The present investigation is initiated by the discovery onboard the {\itshape Kepler} spacecraft of the powerful optical radiation in the course of non-stationary processes on low-mass late-type stars (Maehara et~al.~2012, Shibayama et~al.~2013). How do these stars differ from the vast analogous population of the Galaxy? To clarify this issue one should compare the X-ray emission of these stars. Unfortunately so far the X-rays from quiet stellar coronae are nearly undetectable at the distance of about 100 parsec where the stars with superflares reside. However there exists a certain interrelation between active processes in the photosphere and corona. It is known that the soft X-ray radiation correlates with the spotted surface area. The coronae of the stars in question consist of hot dense structures while only 10\% of their surface is covered by spots. In this situation one may use (at some level of certainty) the data on the spottedness as an estimate of the activity characteristics. As an example consider a restricted set of stars observed by {\itshape Kepler} with short cadence (1 min resolution), on which superflares were detected. We use the data on 61 such stars. The stars with superflares are taken from the list of Balona~(2015). The photospheric activity of these stars is manifested in the variability of the optical and NIR radiation registered by {\itshape Kepler}. That's why we first compared the amplitudes of the variability with the rotation periods of these stars. In the previous works these amplitudes of rotational modulation (ARM) are denoted $\Delta F / F$ or $R_\mathrm{per}$. They are determined by the method of Basri et~al.~(2011) modified by McQuillan et~al.~(2013). Physically these amplitudes are the lower estimates of the fraction of the spotted surface of the star, commonly expressed in parts per million (ppm). In fact the actual area of the spots is larger than that derived from ARM since spots can be distributed in different ways over the stellar surface (e. g., only on the active longitudes or uniformly across all the longitudes). Moreover, some stars are observed from the pole and their ARM is obviously close to zero regardless of the spottedness (on the timescale of the rotation period).

However we can compare the average ARM of all the stars in the catalogue of McQuillan et~al.~(2014) with that of the stars with superflares from the catalogue of Balona~(2015). One can expect that if the larger spottedness is inherent to superflare stars then these stars will demonstrate larger ARM as well. Thus, we can plot the ARM distribution of the stars in the catalogue of McQuillan et~al.~(2014) (which is already done in their article for several ranges of the effective temperature) and compare it with the ARM values observed on the superflare stars.

\begin{figure}
\includegraphics[scale=0.8]{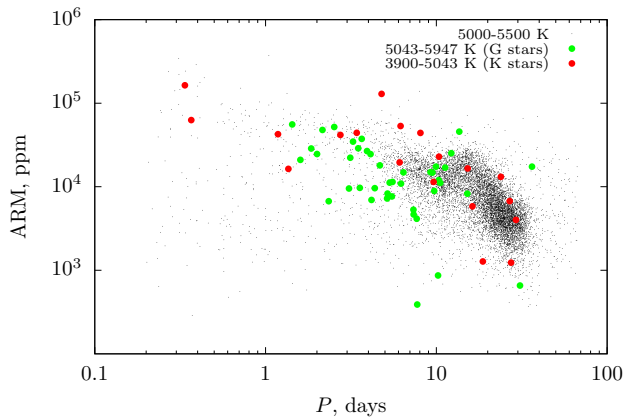}
\caption{\label{fig3} Stars from the catalogue of McQuillan et~al.~(2014) in the $T_\mathrm{eff}$ range 5000 -- 5500 K (black dots); stars from the catalogue of Balona~(2015) (green circles: G stars; red circles: K stars).}
\end{figure}

In Fig.~\ref{fig3}
we show the ARM values for late G stars from the catalogue of McQuillan et~al.~(2014) and those 61 GK stars on which superflares were detected (in his paper Balona~2015 reports on 209 flare stars, but only 61 of them are GK stars and also found in McQuillan~et~al.~2014). One can see that the superflare stars fall mainly in the region where stars with small periods and high variability are located. It is worth noting that several K stars fell into the region with larger periods and smaller variability. The bimodal character of the rotation period distribution of the stars in the temperature range 5000 -- 5500 K is still pronounced while for cooler stars it is manifested more distinctly. The division of the stars in two subgroups is traced in the values of ARM and $P$. This reflects the simultaneous existence of stars with the saturated activity and with the solar-type activity both within a given range of rotation periods (say, 5 -- 15 days) and within a restricted range of $T_\mathrm{eff}$. The bimodality is pronounced in X-rays as well as in $\mathrm{H}_\alpha$, where two levels of the flux for K stars clearly exist (see Fig.~7 in Mart\'{\i}nez-Arn{\'a}iz et~al.~2011). Thus, G and K superflare stars are predominantly fast rotating young objects, but some of them belong to the stars with the solar-type activity. Note that these results agree with the detailed statistical analysis of Shibayama et~al.~(2013) who show that the majority of superflare stars rotate with periods of less than 10 days. Some of these stars rotate with a period of more than 15 days.

\section{Results and discussion}
So far the consideration of the dependence of the coronal activity index on the rotation period was performed on the basis of the catalogue created by Wright et~al.~(2011). Reiners et~al.~(2014) proposed a more adequate method for describing the obtained relation; this method allowed to analyse the behaviour of $\lg L_\mathrm{X} / L_\mathrm{bol}$ in more detail. However this analysis incorporated the dataset of low-mass late-type stars as a whole. In order to obtain new information we have developed the ideas of Reiners et~al.~(2014). This allowed us to investigate the aforementioned relation separately for three spectral types using the same data.

The method we used allowed us to find the rotation period at which the activity mode changes from the saturation to the solar type. For the G2, K4 and M3 stars these periods appeared to be 1.1, 3.3 and 7.2 days respectively. This means that the coronae of red dwarfs longer (up to the periods of about 10 days) exist in the saturation mode. Warmer stars exit the saturation mode earlier. In other words the coronae of red dwarfs at the period of 10 days still remain dense and hot which can be related to the predominant role of local magnetic fields on M stars. This can serve as an argument in favour of a known mechanism of the heating of the red dwarfs coronae due to numerous microflares.

The investigation presented here points that in the rotation period range from several to approximately 20 days there exist stars with both modes of activity. This is beyond the frame of the one-parametric gyrochronology and points at the existence of additional factors which affect the activity level. First of all the difference between the activity levels in the chromosphere derived from $\mathrm{H}_\alpha$ line analysis and in the corona are apparently caused by different spottedness. On the Sun such difference appears when one or several large spots emerge in the active region. In open clusters with more or less confident age estimation stars which have already braked and a number of still fast rotating stars are present. These two populations are studied in detail by Barnes~(2003) (see Fig.~2 therein) who revealed their features. The most of the stars in the cluster demonstrate monotonic growth of the rotation period from F to M stars and subsequently settle at a constant period corresponding to the gyroage of the cluster. The rotation periods of the second (slowly rotating) cluster population with the age of tens of million years gradually decrease from G to M stars. The fraction of such fast rotating stars abruptly falls down in clusters older than 200 Myr.

The difference between the rotation periods corresponding to the transition from one activity mode to another among stars of different spectral types can be related to the difference in mass and consequently the internal structure of these stars. First of all this applies to the convective zone. The depth of the convective zone increases from F to M dwarfs, and this changes the operation of the dynamo mechanism. One can suppose that the disclosed dependence of the point of change of the activity mode on the spectral type is associated with the gradual change of the role of large and small scale magnetic fields in the formation of the activity.

Let us make a remark on one of the results concerning the stars with fundamental parameters close to those of the Sun. We have shown that G stars change the activity mode at the rotation period slightly longer than 1 day. In fact the activity of such a G star substantially exceeds all the acceptable values for the modern Sun at the highest maximum of the cycle. We use the conventional name ``young Sun'' as applied to a G dwarf with the rotation period of about 10 days (or with the age of 1 Gyr) which can already possess the activity of solar type with an established regular cycle. There pass of order 500 Myr from the saturated mode of activity until the solar-type activity sets. It is these stars with the rotation periods from 1 to 10 days where non-stationary processes with the energy and character unconceivable on the modern Sun can take place.

We have analysed the location of the superflare stars on the ARM -- rotation period diagram relative to a large sample of stars observed by {\itshape Kepler} and have concluded that G and K superflare stars are mainly fast rotating young objects, but some of them belong to the stars with the solar type of activity.

It is worth noting that the data we have used in Section~\ref{main} are a compilation of many researches (see Wright et~al.~2011). Unfortunately, we have little data, especially for M dwarfs with the solar-type activity. Obviously the emergence of vast homogenious data on the X-ray luminosity of low-mass stars will allow to come to more definitive conclusions on the problem under discussion.

\begin{acknowledgments}
This work was supported by Russian Foundation for Basic Research (grants 14-02-00922 and 15-02-06271) and Programme of support of leading scientific schools (grant NSh-9670.2016.2). 
\end{acknowledgments}

\appendix*

\section{Bootstrap method}
\begin{figure}
\includegraphics[scale=0.55]{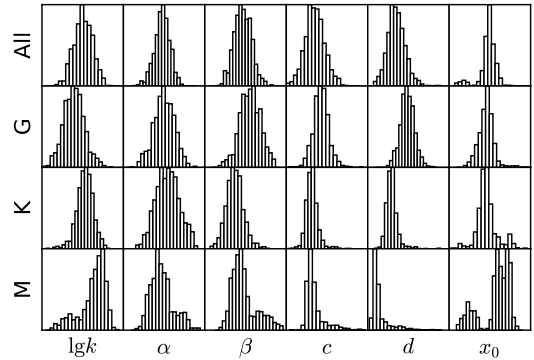}
\caption{\label{fig4} The distribution of the parameters of the dependency $\lg L_\mathrm{X} / L_\mathrm{bol}(\lg R, \lg P)$ derived from the synthetic datasets.}
\end{figure}

Assume that for a number of variables there is a set of $n$ measurements (or observations). For instance, in our case for the variables $\lg L_\mathrm{X} / L_\mathrm{bol}$, $\lg R$, $\lg P$ there are 824 observations. The goal is to find out from these data a functional dependency between these variables, in our case $\lg L_\mathrm{X} / L_\mathrm{bol} = f(\lg R, \lg P)$. This dependency is characterised by several parameters ($k$, $\alpha$, $\beta$, $c$, $d$, $x_0$). Due to the errors in the measurements or observations these parameters are also determined with some errors. Bootstrap is one of the methods of these errors determination (see, e. g., Press et~al.~2007). It is attractive due to its simplicity and consists in the following. Call the whole set of the observations the actual dataset. Now generate a synthetic dataset according to the following procedure. As the first observation in the synthetic dataset choose a random observation from the actual dataset. The chosen observation is then ``returned'' to the actual dataset. In the same way choose another observation from the actual dataset and add it to the synthetic dataset (with the probability of $1/n$ it will be one and the same observation since each observation --- this is essential --- is ``returned'' to the actual dataset after having been chosen). Repeating this procedure $n$ times construct a synthetic dataset with $n$ observations. Note that almost certainly some observations from the actual dataset will be absent in this synthetic dataset while some others will be duplicated (possibly several times). The more synthetic datasets one constructs the better. In our problem we have constructed 1000 sets. In terms of combinatorics the synthetic datasets are no more than the combinations with repetitions. Their total number  $\left(\!{n \choose n}\!\right) = (2n-1)!/n!(n-1)!$ is huge so that the synthetic sets are almost definitely different. Further for each synthetic set one computes the sought-for parameters in the usual way and finally one can plot the distribution of each parameter and find the mean squared deviation. In Fig.~\ref{fig4}
we show the distributions of all the parameters in four cases: when the whole set of 824 stars was considered as the actual set and when it was restricted to include only G, K or M stars. This figure is only an illustration, the scales of $x$ and $y$ axes vary from one plot to another. However one sees that sometimes the distributions deviate from normality, especially in the case of M stars. But to characterise the accuracy of the derived parameters we still use the mean squared deviation.


\end{document}